# The effect of stripe domain structure on dynamic permeability of thin ferromagnetic films with out-of-plane uniaxial anisotropy

N.A. Buznikov [*], K.N. Rozanov

*Institute for Theoretical and Applied Electrodynamics, Russian Academy of Sciences, Moscow 125412, Russia*

**Abstract**

The permeability is calculated for a thin ferromagnetic film with the stripe domain structure and out-of-plane uniaxial magnetic anisotropy. Analytical expressions for the frequency dependence of components of permeability tensor are derived with the use of the Smit−Beljers method, with the thickness of domain walls and the domain wall motion being neglected. The effect of the domain width and the angle between the anisotropy axis and the film plane on the frequency dependence of the permeability is analyzed. General equations relating the static permeability components and the ferromagnetic resonance frequencies are found. The results of the approach are applied to the derivation of the constraint for the microwave permeability of thin ferromagnetic films. The analysis of the constraint as a function of the axis deviation angle, the domain aspect ratio and the damping parameter allows the conditions to be found for maximal microwave permeability. The results obtained may be useful in connection with the problem of developing high-permeable microwave magnetic materials.

*PACS:* 75.70.-i; 76.50.+g

*Keywords:* Thin films − ferromagnetic; Microwave; Permeability; Ferromagnetic resonance

---

[*] Corresponding author. *E-mail address:* n_buznikov@mail.ru

# 1. Introduction

An increased interest has been given recently to the microwave performance of thin ferromagnetic films. The interest is rooted on possible applications of such films as micro-inductors, micro-transformers, magnetic sensors, magneto-optical media, tunable microwave filters, magneto-dipole antennas, etc. [1–5] For these applications, high values of microwave permeability are needed.

The frequency dependence of microwave permeability is typically due to the ferromagnetic resonance. In many magnetic materials, the permeability varies slightly with frequency until the resonance appears, where an abrupt drop in the permeability is observed. Above the resonance, the permeability takes low values that are inappropriate for most applications. Therefore, the opportunity for microwave applications of a magnetic material can be estimated from its static permeability $\mu(0)$ and resonance frequency $\omega_r$. The static permeability characterizes the magnetic performance in the operating frequency band. The resonance frequency is an estimate for the cutoff frequency, above which the material is no more permeable enough and magnetic loss increases greatly. Therefore, in a material exhibiting high values of the microwave permeability, both $\omega_r$ and $\mu(0)$ should be as high as possible.

Typically, these values reveal a complicated dependence on the composition of the material, its magnetic and crystallographic structure, peculiarities of manufacturing process, etc. However, the resonance frequency and the static permeability are related each to other: an increase of $\mu(0)$ leads to a decrease in $\omega_r$, and vise versa. In a bulk polycrystalline magnetic material, this relation is given by Snoek's law [6]:

$$[\mu(0) - 1]\omega_r = (2/3)(\gamma 4\pi M_s), \tag{1}$$

where $\gamma$ is the gyromagnetic constant and $M_s$ is the saturation magnetization. The right-hand part of Eq. (1) depends on the saturation magnetization only and does not involve structure-dependent values, such as the anisotropy field. Therefore, it is determined by the composition of the material rather than by peculiarities of its structure and preparation process. Snoek's law allows a useful estimate for the limiting value of microwave permeability of a bulk magnetic material to be obtained based on its saturation magnetization.

For thin ferromagnetic films, Snoek's law is not valid, and the relation between $\omega_r$ and $\mu(0)$ has the form [7,8]

$$[\mu(0) - 1]\omega_r^2 = (\gamma 4\pi M_s)^2 (1 + H_a / 4\pi M_s), \qquad (2)$$

where $H_a$ is the anisotropy field. Since $H_a / 4\pi M_s \ll 1$ in high-permeability films, the right-hand part of Eq. (2) depends on the saturation magnetization only, the same as that in Eq. (1). Eq. (2) provides a constraint for the microwave permeability of thin magnetic films that is analogous to Snoek's law for bulk magnetic materials.

It follows from Eqs. (1) and (2) that if $\omega_r < \gamma 4\pi M_s$, Eq. (2) produces higher values of the static permeability and, therefore, higher microwave permeability than Eq. (1) does. For typical high-permeability films, the value of $\gamma 4\pi M_s$ is several GHz or several tens GHz. It is for this reason that thin ferromagnetic films are considered to be promising materials for microwave applications.

Eq. (2) has been obtained for a single-domain infinite thin film with uniaxial in-plane anisotropy. It is of interest to obtain similar constraints for more realistic models of ferromagnetic films. It has been shown recently that the limiting value of the microwave permeability in films of finite thickness is less than that predicted by Eq. (2) and depends on the ratio of the film thickness to the film transverse dimension [9]. The effect of other factors that influence on magnetic properties of films, such as the existence of a stripe domain structure and the deviation of the anisotropy axis from the film plane, has not been considered previously.

The effect of a domain structure on the ferromagnetic resonance has been studied in detail. In a film with the domain structure, two modes of the ferromagnetic resonance appear, that is known as the Polder−Smit effect [10]. These modes occur due to different orientations of the magnetization in two types of domains. The low-frequency mode is excited by the component of the alternating magnetic field that is transverse to the domain walls. In this case, there are no dynamic poles at the domain walls. The high-frequency mode is excited when the alternating magnetic field is parallel to the domain walls. For this mode, the alternating magnetization vectors in the neighboring domains induce dynamic poles at the domain walls. A more general theory of the ferromagnetic resonance in thin films with the account for a bias magnetic field has been proposed in Refs. [11,12] under the assumption of thin domains. More recently, equations for the resonance frequencies have been improved by taking into account a finite domain width [13,14]. Although the calculation technique is well established for the high-frequency response of thin films for the case of the stripe domain structure, most papers deal with only resonance frequencies arising in this problem. The exception is Ref. [15], where the permeability has been found for the multi-domain film with the anisotropy axis perpendicular to the film plane.

The anisotropy axis in a soft magnetic film may have an arbitrary angle with the film plane [16–19]. Again, most studies deal with the calculation of the resonance frequencies leaving aside the problem of the microwave permeability. Therefore, it is of importance to study the effect of the out-of-plane anisotropy on the microwave permeability of a thin film. Preliminary results have been obtained in Ref. [20], where a single-domain film has been considered.

This paper deals with ferromagnetic films having a stripe domain structure and an arbitrary angle between the anisotropy axis and the film plane. Analytical expressions for the frequency dependence of components of the permeability tensor are derived with the use of the Smit–Beljers method [11]. Using expressions for the permeability components, the constraint for the microwave permeability is found. The obtained constraint for the microwave permeability of a ferromagnetic film with the stripe domain structure and out-of-plane anisotropy has a more complicated form than Eqs. (1) and (2) and involves structure-dependent parameters. The analysis of the constraint as a function of the axis deviation angle, the domain aspect ratio and the damping parameter allows one to find the conditions for maximal microwave permeability.

## 2. Permeability tensor of thin film

### 2.1. Basic equations

The geometry of the problem under study is shown in Fig. 1. The film of thickness $d$ is in $x$–$y$ plane, and the anisotropy axis lies in $y$–$z$ plane. The angle between the anisotropy axis and the film plane is $\psi$, $0 < \psi < \pi/2$. The stripe domain structure is supposed to include equally spaced parallel domains of the width $a$. It is assumed that the domain walls are transverse to $x$-axis and do not move. We assume also that the film is infinite along $x$- and $y$-axes.

The approach to calculate the permeability tensor of the sample with the stripe domain structure is well known [21,22]. The response of magnetization to the alternating magnetic field $\mathbf{h}\exp(-i\omega t)$ is governed by the Landau–Lifshitz–Gilbert equation

$$\partial \mathbf{M}_j / \partial t = -\gamma(\mathbf{M}_j \times \mathbf{H}_{\mathrm{ef},j}) + \alpha(\mathbf{M}_j \times \partial \mathbf{M}_j / \partial t) M_\mathrm{s}^{-1}. \qquad (3)$$

Here $j = 1, 2$ are the indexes corresponding to two opposite orientations of the magnetization in neighboring domains and $\alpha$ is the Gilbert damping parameter. The effective magnetic field $\mathbf{H}_{\mathrm{ef},j}$ is given by

$$\mathbf{H}_{\mathrm{ef},j} = -\delta U / \delta \mathbf{M}_j + \mathbf{h}\exp(-i\omega t), \qquad (4)$$

where $U$ is the free energy density. It consists of two contributions, the anisotropy energy, $U_a$, and the demagnetizing energy associated with the domain structure, $U_m$: $U = U_a + U_m$. The anisotropy term $U_a$ can be presented in the following form [22]:

$$U_a = -[(\mathbf{H}_a \mathbf{M}_1)^2 + (\mathbf{H}_a \mathbf{M}_2)^2] / 4 H_a M_\mathrm{s}. \qquad (5)$$

If the domains are oriented along $y$-axis, the demagnetizing energy $U_m$ is given by [13–15]

$$U_m = (\pi/2)(M_{1z} + M_{2z})^2 + (N_{zz}/8)(M_{1z} - M_{2z})^2 + (N_{xx}/8)(M_{1x} - M_{2x})^2, \qquad (6)$$

where $M_{jx}$, $M_{jz}$ are $x$- and $z$-components of the magnetization in the domains, respectively. The first term in the right-hand side of Eq. (6) corresponds to the demagnetizing energy associated with the average magnetization component, which is normal to the film plane. Since the domains are assumed to be infinite along $y$-axis and therefore $N_{yy} = 0$, the demagnetizing energy attributed to the domain structure is specified by the demagnetizing factors $N_{xx}$ and $N_{zz}$ that can be found as [15,23,24]

$$N_{zz} = \frac{64a}{\pi^2 d} \times \sum_{n=1}^{\infty} \frac{\sin^2(\pi n/2)}{n^3} \times \frac{\sinh(nq)}{\sinh(nq) + [1 + Q^{-1}]^{1/2} \cosh(nq)}, \qquad (7)$$

$$N_{xx} = 4\pi - N_{zz},$$

where $q = (\pi d / 2a)[1 + Q^{-1}]^{1/2}$; $Q = \omega_a / \omega_m$ is the film quality factor; $\omega_m = \gamma 4\pi M_s$ and $\omega_a = \gamma H_a$ are the characteristic frequencies associated with the saturation magnetization and the anisotropy field, respectively. The quality factor $Q$ is an important parameter of magnetic materials that determine the magnetic performance. Although Eq. (7) has been derived for the films with high quality factor $Q$, it is known to be valid in the case of low $Q$ as well [24].

It is assumed further that the micromagnetic configuration is always the stripe domain structure. This approximation is valid if the exchange energy is negligible. In soft magnetic films, $Q \ll 1$, the stray-field-free domain configuration can occur. In this case, the analysis of the stability of various magnetic structures in the films can be made by solving numerically the complete set of micromagnetic equations [25–27] and by approximate analytical methods [28,29]. The results of calculations made by means of the analytical model based on the variation of a one-dimensional magnetization show that there is a critical line in the phase diagram $Q$–$d$ separating the stripe domain structure and the uniform magnetization configuration in films with perpendicular anisotropy [29]. This means that there is no stripe domain structure in very thin soft magnetic films.

It follows from Eq. (7) that the quality factor $Q$ effects significantly on the demagnetizing factors. The physical reason for this dependence is as follows. For magnetic materials with very high anisotropy field, the equilibrium magnetization is along the anisotropy field. If the anisotropy field is not too high, the stray fields aspire to rotate the magnetization vectors away from the anisotropy field direction. This mechanism for the $Q$-dependence of the demagnetizing factors has been suggested in Ref. [30]. The influence of the quality factor on the demagnetizing factors in periodic domain structures has been taken into account firstly in Ref. [31]. It should be noted that in real magnetic materials the quality factor $Q$ and the equilibrium domain aspect ratio $a/d$ are related to each other [32]. Since the detailed micromagnetic analysis of the possible magnetic structures is beyond the scope of this paper, we assume for simplicity that these factors may vary independently. Indeed, before using of results obtained further the appearance of the domain structure with given aspect ratio should be checked.

Approximate analytical expressions for the demagnetizing factor $N_{zz}$ can be obtained from Eq. (7) in the case of wide domains, $a/d \gg 1$:

$$N_{zz} \cong 4\pi, \quad a/d \gg 1 \tag{8}$$

and in the case of narrow domains, $a/d \ll 1$:

$$N_{zz} \cong \frac{56\zeta(3)}{\pi^2} \times \frac{a/d}{1+(1+Q^{-1})^{1/2}}, \quad a/d \ll 1, \tag{9}$$

where $\zeta(3) \approx 1.20$ is the Riemann $\zeta$ function of 3.

## 2.2. Equilibrium magnetization

The static equilibrium values of the magnetization vectors in the domains, $\mathbf{M}_1^0$ and $\mathbf{M}_2^0$ and the effective magnetic field $\mathbf{H}_{\text{ef},j}^0$ are governed by equation

$$\mathbf{M}_j^0 \times \mathbf{H}_{\text{ef},j}^0 = 0. \tag{10}$$

Assuming $\mathbf{h} = 0$, we obtain from Eqs. (4)–(7) and (10) that

$$\begin{aligned}
M_{1x}^0 &= M_{2x}^0 = 0, \\
M_{1y}^0 &= -M_{2y}^0 = M_s \cos\theta, \\
M_{1z}^0 &= -M_{2z}^0 = M_s \sin\theta, \\
H_{\text{ef},1x}^0 &= H_{\text{ef},2x}^0 = 0, \\
H_{\text{ef},1y}^0 &= -H_{\text{ef},2y}^0 = (H_a/2)\cos\psi \cos(\psi-\theta), \\
H_{\text{ef},1z}^0 &= -H_{\text{ef},2z}^0 = (H_a/2)\sin\psi \cos(\psi-\theta) - (N_{zz}M_s/2)\sin\theta,
\end{aligned} \tag{11}$$

where the angle $\theta$ between the magnetization and $y$-axis is given by

$$\tan 2\theta = \frac{P \sin 2\psi}{1 + P \cos 2\psi}, \quad P = 4\pi Q/N_{zz}. \tag{12}$$

The equilibrium angle $\theta$ is shown in Fig. 2 as a function of the anisotropy axis deviation angle $\psi$ at different values of the factor $P$. It follows from Fig. 2 that function $\theta(\psi)$ differs significantly for the cases $P > 1$ and $P < 1$. If $P > 1$, the equilibrium angle $\theta$ increases monotonically with $\psi$, whereas if $P < 1$, the angle $\theta$ has a peak and then decreases with an increase of the anisotropy axis deviation angle $\psi$. For soft magnetic films with wide domains, $P \ll 1$, the equilibrium angle $\theta$ is small, hence the magnetization vector deviates slightly from the film plane. In this case, Eq. (12) is reduced to

$$\theta = (2\pi Q / N_{zz}) \sin 2\psi [1 - (4\pi Q / N_{zz}) \cos 2\psi]. \tag{13}$$

*2.3. Permeability tensor*

The dynamic response of the film to a weak alternating magnetic field is analyzed by considering small oscillations of the magnetization vectors in the domains about the equilibrium state. Let an alternating magnetic field have three components, and the magnetization and the effective magnetic field perturbed by the alternating magnetic field $\mathbf{h}\exp(-i\omega t)$ be given by

$$\begin{aligned}\mathbf{M}_j &= \mathbf{M}_j^0 + \mathbf{m}_j \exp(-i\omega t), \\ \mathbf{H}_{\text{ef},j} &= \mathbf{H}_{\text{ef},j}^0 + \mathbf{h}_{\text{ef},j} \exp(-i\omega t).\end{aligned} \tag{14}$$

The linearized Landau–Lifshitz–Gilbert equation has the form

$$i\omega \mathbf{m}_j = \gamma(\mathbf{M}_j^0 \times \mathbf{h}_{\text{ef},j} + \mathbf{m}_j \times \mathbf{H}_{\text{ef},j}^0) + i\alpha\omega(\mathbf{M}_j^0 \times \mathbf{m}_j)M_\text{s}^{-1}, \tag{15}$$

where the components of $\mathbf{h}_{\text{ef},j}$ are written as

$$\begin{aligned}h_{\text{ef},1x} &= h_x/2 - (N_{xx}/4)(m_{1x} - m_{2x}), \\ h_{\text{ef},2x} &= h_x/2 + (N_{xx}/4)(m_{1x} - m_{2x}), \\ h_{\text{ef},1y} &= h_y/2 + (H_\text{a}/2M_\text{s})(m_{1y}\cos\psi + m_{1z}\sin\psi)\cos\psi, \\ h_{\text{ef},2y} &= h_y/2 + (H_\text{a}/2M_\text{s})(m_{2y}\cos\psi + m_{2z}\sin\psi)\cos\psi, \\ h_{\text{ef},1z} &= h_z/2 + (H_\text{a}/2M_\text{s})(m_{1y}\cos\psi + m_{1z}\sin\psi)\sin\psi \\ &\quad - \pi(m_{1z} + m_{2z}) - (N_{zz}/4)(m_{1z} - m_{2z}), \\ h_{\text{ef},2z} &= h_z/2 + (H_\text{a}/2M_\text{s})(m_{2y}\cos\psi + m_{2z}\sin\psi)\sin\psi \\ &\quad - \pi(m_{1z} + m_{2z}) + (N_{zz}/4)(m_{1z} - m_{2z}).\end{aligned} \tag{16}$$

From Eqs. (11), (15) and (16), the susceptibility tensor $\hat{\chi}$ of the film that couples perturbations of the magnetization and the alternating magnetic field, $(\mathbf{m}_1 + \mathbf{m}_2)/2 = \hat{\chi}\mathbf{h}$, can be derived. After that, the components $\mu_{jk}$ of the permeability tensor $\hat{\mu}$ are calculated as $\hat{\mu} = \hat{I} + 4\pi\hat{\chi}$, where $\hat{I}$ is the unit tensor. This yields

$$\begin{aligned}\mu_{xx} &= 1 + \omega_\text{m}^2[(N_{zz}/4\pi)\cos 2\theta + Q\cos\{2(\psi-\theta)\} - i\alpha\omega/\omega_\text{m}]/\Omega_1^2, \\ \mu_{yy} &= 1 + \omega_\text{m}^2 \sin\theta[(1 - N_{zz}/2\pi)\sin\theta + Q\sin\psi\cos(\psi-\theta) - i\alpha\omega\sin\theta/\omega_\text{m}]/\Omega_2^2, \\ \mu_{yz} &= \mu_{zy} = -\omega_\text{m}^2 \sin\theta[(1 - N_{zz}/4\pi)\cos\theta + Q\cos\psi\cos(\psi-\theta) - i\alpha\omega\cos\theta/\omega_\text{m}]/\Omega_2^2, \\ \mu_{zz} &= 1 + \omega_\text{m}^2 \cos\theta[(1 - N_{zz}/4\pi)\cos\theta + Q\cos\psi\cos(\psi-\theta) - i\alpha\omega\cos\theta/\omega_\text{m}]/\Omega_2^2, \\ \mu_{xy} &= \mu_{yx} = \mu_{xz} = \mu_{zx} = 0.\end{aligned} \tag{17}$$

In Eq. (17) we denote

$$\Omega_1^2 = (1+\alpha^2)(\omega_1^2 - \omega^2) - i\alpha\omega\omega_m[(N_{zz}/4\pi)(\cos^2\theta - 2\sin^2\theta)$$
$$+ Q\{2\cos^2(\psi-\theta) - \sin^2(\psi-\theta)\}],$$
$$\Omega_2^2 = (1+\alpha^2)(\omega_2^2 - \omega^2) - i\alpha\omega\omega_m[(2-N_{zz}/4\pi)\cos^2\theta + (1-3N_{zz}/4\pi)\sin^2\theta$$
$$+ Q\{2\cos^2(\psi-\theta) - \sin^2(\psi-\theta)\}],$$
(18)

The ferromagnetic resonance frequencies $\omega_1$ and $\omega_2$ are given by

$$\omega_1^2 = \omega_m^2[(N_{zz}/4\pi)^2 \sin^2\theta + Q(N_{zz}/4\pi)\{\cos^2\psi - 2\sin\psi\sin\theta\cos(\psi-\theta)\}$$
$$+ Q^2\cos^2(\psi-\theta)]/(1+\alpha^2),$$
$$\omega_2^2 = \omega_m^2[2(N_{zz}/4\pi)^2 \sin^2\theta - N_{zz}/4\pi + \cos^2\theta$$
$$+ Q\{F(\psi,\theta) - (N_{zz}/4\pi)G(\psi,\theta)\} + Q^2\cos^2(\psi-\theta)]/(1+\alpha^2),$$
$$F(\psi,\theta) = \cos\{2(\psi-\theta)\} + \cos\psi\cos\theta\cos(\psi-\theta),$$
$$G(\psi,\theta) = \cos\{2(\psi-\theta)\} + \sin\theta[\cos\psi\sin(\psi-\theta) + 2\sin\psi\cos(\psi-\theta)].$$
(19)

Eq. (17) allows for the Polder-Smit effect: the permeability component $\mu_{xx}$ has the resonance frequency $\omega_1$, and all other components have different resonance frequency $\omega_2$. Note that at $Q > 1$ and $\psi = \theta = \pi/2$, Eq. (19) coincides with the result obtained for the resonance frequencies of thin ferromagnetic film having the perpendicular anisotropy [15].

*2.4. Comparison with single-domain film*

To understand the effect of the domain structure on the permeability, it is helpful to compare the permeability tensor (17) with that obtained for a single-domain film. The single-domain case can be treated by the same approach as above with $\mathbf{M}_1 = \mathbf{M}_2$ and $N_{zz} = 4\pi$, that yields [20]

$$\mu_{xx} = 1 + \omega_m^2[\cos 2\theta + Q\cos\{2(\psi-\theta)\} - i\alpha\omega/\omega_m]/\Omega_0^2,$$
$$\mu_{xy} = -\mu_{yx} = -i\omega\omega_m \sin\theta/\Omega_0^2,$$
$$\mu_{xz} = -\mu_{zx} = i\omega\omega_m \cos\theta/\Omega_0^2,$$
$$\mu_{yy} = 1 + \omega_m^2 \sin\theta[-\sin\theta + Q\sin\psi\cos(\psi-\theta) - i\alpha\omega\sin\theta/\omega_m]/\Omega_0^2,$$
$$\mu_{yz} = \mu_{zy} = -\omega_m^2 \sin\theta[Q\cos\psi\cos(\psi-\theta) - i\alpha\omega\cos\theta/\omega_m]/\Omega_0^2,$$
$$\mu_{zz} = 1 + \omega_m^2 \cos\theta[Q\cos\psi\cos(\psi-\theta) - i\alpha\omega\cos\theta/\omega_m]/\Omega_0^2,$$
(20)

where

$$\Omega_0^2 = (1+\alpha^2)(\omega_0^2 - \omega^2) - i\alpha\omega\omega_m[\cos^2\theta - 2\sin^2\theta$$
$$+ Q\{2\cos^2(\psi-\theta) - \sin^2(\psi-\theta)\}],$$
$$\omega_0^2 = \omega_m^2[\sin^2\theta + Q\{\cos^2\psi - 2\sin\psi\sin\theta\cos(\psi-\theta)\}$$
$$+ Q^2\cos^2(\psi-\theta)]/(1+\alpha^2).$$
(21)

The expressions for the diagonal components of the permeability tensor and for $\mu_{yz}$ in Eq. (20) can be derived immediately from Eq. (17), assuming that the domains are wide, $a/d \gg 1$. In this case, $N_{zz} \approx 4\pi$ and the demagnetizing fields do not affect the magnetic performance of the film. Two modes of ferromagnetic resonance in the stripe-domain structure merge into a single peak, $\omega_1 = \omega_2 = \omega_0$. Correspondingly, the diagonal components of the permeability take values typical for a single-domain film.

The off-diagonal components of the permeability tensor are beyond the scope of the further discussion. Notice, however, that the $\mu_{xy}$ and $\mu_{xz}$ components are non-zero for a single-domain film. On the contrary, in the case of a stripe domain structure, these components are zero. The difference is attributed to the non-zero static magnetization of a single-domain film that can not appear in a stripe domain structure.

## 3. Analysis

It follows from Eqs. (17) and (18) that the components of the permeability tensor of a film with a stripe domain structure depend on four parameters, namely, the anisotropy axis deviation angle $\psi$, the quality factor $Q$, the domain aspect ratio $a/d$ and the damping parameter $\alpha$. The effect of these parameters on the ferromagnetic resonance frequencies and the permeability components is considered below. Further, we draw attention to the case of soft magnetic films, $Q \ll 1$, that is of importance for applications.

### 3.1. Resonance frequencies

Fig. 3 shows the resonance frequencies $\omega_1$ and $\omega_2$ calculated from Eq. (19) as a function of the domain aspect ratio $a/d$ at $\psi = \pi/4$ and different values of the quality factor $Q$. The resonance frequency $\omega_1$ increases monotonically with $a/d$, from $\omega_1 \cong \omega_a/(1+\alpha^2)^{1/2}$ at $a \ll d$ to $\omega_1 \cong \omega_0$ at $a \gg d$. For narrow domains, the resonance frequency $\omega_2$ is much higher than $\omega_1$. The value of $\omega_2$ increases with the domain aspect ratio, achieves a peak $\omega_{2\max} \cong \omega_m/(1+\alpha^2)^{1/2}$ at some value of $a/d$ and then drops sharply. At high values of the domain aspect ratio, the frequencies $\omega_1$ and $\omega_2$ coincide and are equal to the resonance frequency $\omega_0$ of a single-domain film.

Shown in Fig. 4 are the dependences of $\omega_1$ and $\omega_2$ on the anisotropy axis deviation angle $\psi$ at $Q = 10^{-2}$ and different aspect ratio $a/d$. It is seen from Fig. 4 that the resonance frequency $\omega_1$ decreases with an increase of the axis deviation angle $\psi$ for wide domains, whereas in the case of narrow domains, $a/d \ll 1$, the value of $\omega_1$ is almost independent of $\psi$. On the contrary, the dependence of $\omega_2$ on $\psi$ is slight for wide domains and becomes more pronounced at $a/d \ll 1$. It follows from Fig. 4 that $\omega_2$ decreases more and more sharply with a growth of the axis deviation angle for narrow domains.

For soft magnetic films expressions for the resonance frequencies can be simplified. In the case of narrow domains, $a/d \ll 1$, taking into account that $\theta \approx \psi$, we have:

$$\begin{aligned}
\omega_1^2 &= \omega_m^2 Q[(N_{zz}/4\pi)\cos^2\psi + Q]/(1+\alpha^2), \\
\omega_2^2 &= \omega_m^2[(1 - N_{zz}/4\pi)\cos^2\psi + Q(1 + \cos^2\psi - N_{zz}/4\pi) + Q^2].
\end{aligned} \quad (22)$$

For wide domains, $a/d \gg 1$, using Eq. (13), we find for the resonance frequencies from Eq. (19)

$$\begin{aligned}
\omega_1^2 &= \omega_m^2 Q \cos^2\psi [(N_{zz}/4\pi) + Q\cos^2\psi]/(1+\alpha^2), \\
\omega_2^2 &= \omega_m^2 [1 - N_{zz}/4\pi + Q\{\cos^2\psi(2 - N_{zz}/4\pi) - \sin^2\psi(1 - N_{zz}/4\pi)\} \\
&\quad + (4\pi Q \cos\psi / N_{zz})^2 \{\cos^2\psi + (1 - N_{zz}/4\pi)(5N_{zz}\sin^2\psi/4\pi - 1)\}]/(1+\alpha^2).
\end{aligned} \quad (23)$$

*3.2. Static permeability*

The diagonal components of the static permeability,

$$\begin{aligned}
\mu_{xx}(0) &= 1 + (\omega_m/\omega_1)^2[(N_{zz}/4\pi)\cos 2\theta + Q\cos\{2(\psi - \theta)\}]/(1+\alpha^2), \\
\mu_{yy}(0) &= 1 + (\omega_m/\omega_2)^2 \sin\theta[(1 - N_{zz}/2\pi)\sin\theta + Q\sin\psi\cos(\psi - \theta)]/(1+\alpha^2), \\
\mu_{zz}(0) &= 1 + (\omega_m/\omega_2)^2 \cos\theta[(1 - N_{zz}/4\pi)\cos\theta + Q\cos\psi\cos(\psi - \theta)]/(1+\alpha^2)
\end{aligned} \quad (24)$$

are readily found from Eqs. (17) and (18) by setting $\omega = 0$.

It follows from Eq. (24) that for soft magnetic films, $Q \ll 1$, the largest component of the static permeability is $\mu_{xx}(0)$. Other static permeability components are about unity, since the equilibrium magnetization vector angle $\theta$ is small. The further discussion deals with the component $\mu_{xx}$. The following asymptotic expressions for the static permeability $\mu_{xx}(0)$ in the cases of narrow and wide domains can be found from Eq. (24):

$$\begin{aligned}
\mu_{xx}(0) &= Q^{-1} + 1, \quad a/d \ll 1, \\
\mu_{xx}(0) &= Q^{-1}/\cos^2\psi + 1 - 4\pi\tan^2\psi/N_{zz}, \quad a/d \gg 1.
\end{aligned} \quad (25)$$

Note that in the case of the in-plane anisotropy, $\psi=0$, the static permeability $\mu_{xx}(0)$ is independent of the domain width and $\mu_{xx}(0)=Q^{-1}+1$.

*3.3. Limiting value of microwave permeability*

The effect of the anisotropy axis deviation angle and the domain width on the limiting value of the microwave permeability is analyzed by treating the product of static permeability and squared resonance frequency, the same as in the left-hand part of Eq. (2). A combination of Eqs. (19) and (24) yields the following relationship constraining the microwave permeability value $\mu_{xx}$:

$$[\mu_{xx}(0)-1]\omega_1^2 = \omega_m^2[(N_{zz}/4\pi)\cos 2\theta + Q\cos\{2(\psi-\theta)\}]/(1+\alpha^2). \tag{26}$$

Eq. (26) coincides with Eq. (2) at $N_{zz}=4\pi$, $\psi=0$ and $\alpha\ll 1$. The obtained limiting value of the microwave permeability is analyzed below for soft magnetic films, $Q\ll 1$, as a function of the damping parameter $\alpha$, the anisotropy axis deviation angle $\psi$, the quality factor $Q$ and the domain aspect ratio $a/d$.

If $Q\ll 1$, Eq. and (26) is simplified as

$$\begin{aligned}[\mu_{xx}(0)-1]\omega_1^2 &= \omega_m^2[N_{zz}\cos^2\psi + Q]/(1+\alpha^2), \quad a\ll d, \\ [\mu_{xx}(0)-1]\omega_1^2 &= \omega_m^2[N_{zz} + Q\cos(2\psi)]/(1+\alpha^2), \quad a\gg d.\end{aligned} \tag{27}$$

The effect of damping on the limiting value of the microwave permeability is due to the factor $(1+\alpha^2)^{-1}$ involved in the right-hand part of Eq. (26). Therefore, damping results in a decrease of the left-hand part of Eq. (26), and diminishes the limiting value of the microwave permeability. The effect of damping on the microwave permeability can be even more pronounced due to broadening of resonance absorption lines with an increase $\alpha$. Due to broadening, the real part of the permeability begins to decrease at frequencies well-below the resonance, which results in an additional decrease of the microwave permeability. However, the effects of damping are negligible within the scope of the model under treatment, since Eq. (3) is valid only at $\alpha\ll 1$.

It follows from Eq. (26) that the out-of-plane anisotropy diminishes the in-plane microwave permeability. Fig. 5 shows the parameter $[\mu_{xx}(0)-1]\omega_1^2$ as a function of the quality factor $Q$ for a single-domain film, $N_{zz}=4\pi$. The values of the parameter in Fig. 5 are reduced to $\omega_m^2(1+Q)/(1+\alpha^2)$, so that it is equal to unity when Eq. (2) is held. It is seen from Fig. 5 that an increase of $\psi$ leads to a decrease of the parameter $[\mu_{xx}(0)-1]\omega_1^2$ for films having high anisotropy field. However, for soft magnetic films with $Q \ll 1$, the decrease is small.

The dependence of the parameter $[\mu_{xx}(0)-1]\omega_1^2$ on the quality factor $Q$ is shown in Fig. 6 for different domain aspect ratio $a/d$. For wide domains, $a/d \gg 1$, the value of $[\mu_{xx}(0)-1]\omega_1^2$ decreases with an increase of the quality factor, and the dependence is similar to that obtained for a single-domain film. The behaviour of the parameter $[\mu_{xx}(0)-1]\omega_1^2$ changes drastically in the film with not too wide domains. In this case, as shown above (see Fig. 3), the resonance frequency $\omega_1$ drops that results in the decrease of the parameter $[\mu_{xx}(0)-1]\omega_1^2$. The parameter increases monotonically with the quality factor. However, its value is much lower than in the case of wide domains (see Fig. 6).

The physical mechanism of the decrease of the limiting value of the microwave permeability for films with narrow domains can be understood as follows. For narrow domains, the dynamic demagnetizing fields at the domain walls are high that suppresses the variation of magnetization in the film plane. Then, due to the out-of-plane anisotropy, the magnetization variation in the direction transverse to the film plane is more preferable, that leads to the decrease of the in-plane permeability.

## 4. Conclusions

The limiting value of microwave permeability of a magnetic material can be estimated on the base of the saturation magnetization of the material. It has been shown previously [7–9] that the limiting value in thin ferromagnetic films may be higher than that in bulk materials having the same saturation magnetization. This result has been obtained for the in-plane microwave permeability of a single-domain thin film with uniaxial in-plane anisotropy. In the present paper, a more general problem of a film having a stripe domain structure and out-of-plane anisotropy is treated. The study is aimed at checking the opportunities of obtaining high microwave permeability in such films.

It is shown that both damping and deviation of the anisotropy axis from the film plane result in more severe constraint for the microwave permeability than that predicted for thin ferromagnetic films with in-plane anisotropy, though the effect of these factors is sufficiently small in soft magnetic films. Therefore, the constraint on the microwave permeability established in Refs. [7–9] is valid for the problem under study in the case of a single-domain film.

The stripe domain structure affects significantly the limiting value of in-plane microwave permeability due to the influence of the dynamic demagnetizing fields at the domain walls. For wide domains, these fields are low, and the effect of the domain structure on the microwave permeability eliminates. For films with not too wide domains, the dynamic demagnetizing fields result in a sharp decrease of the microwave permeability, and its limiting value may be several times lower than that in the case of a single-domain film.

The above results are obtained by the use of the Smit–Beljers technique. The discussion on the validity of this technique for real magnetic films is beyond the scope of this paper. In many films, measured microwave permeability reveal more complicated pattern of resonance frequencies [27]. This can be attributed to the effect of domain-wall motion and finite thickness of domain walls that are neglected in the above treatment. This complex resonance patterns can be predicted by micromagnetic calculations [27]. Nevertheless, the application of the Smit–Beljers method allows the constraint for microwave permeability to be obtained in an analytical form, that is useful for understanding the properties of real magnetic materials.

**Acknowledgements**

The authors would like to thank Prof. I.T. Iakubov and Prof. A.L. Rakhmanov for fruitful discussions. This work was supported in part by the Russian Federation President Grant No. 1694.2003.2.

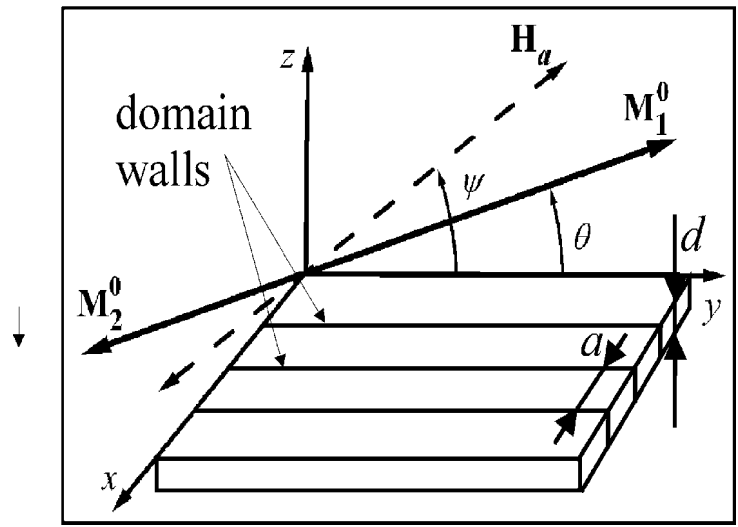

Fig. 1. Geometrical coordinates. The film is in *x–y* plane, anisotropy axis is in *y–z* plane. $\mathbf{M}_1^0$ and $\mathbf{M}_2^0$ are equilibrium magnetization vectors in domains.

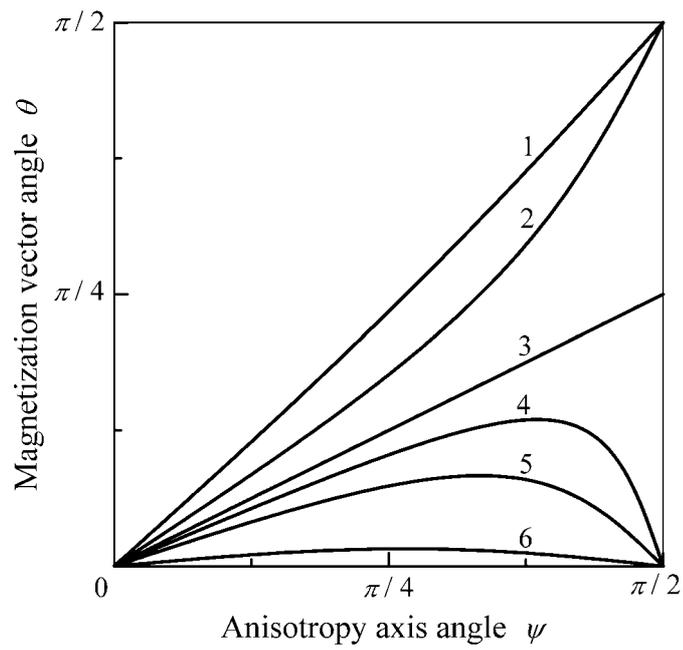

Fig. 2. Equilibrium magnetization vector angle $\theta$ as a function of anisotropy axis deviation angle $\psi$ at different values of factor $P$: 1, $P=10$; 2, $P=2$; 3, $P=1$; 4, $P=0.75$; 5, $P=0.5$; 6, $P=0.1$.

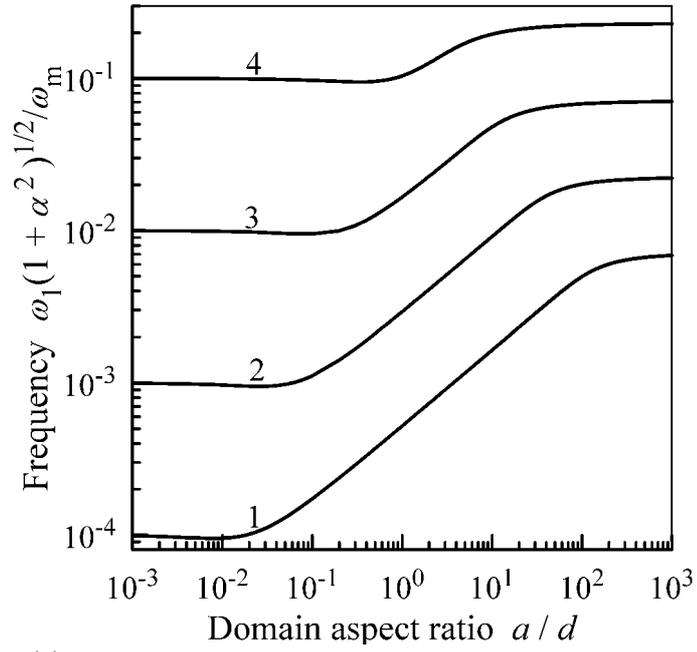

(a)

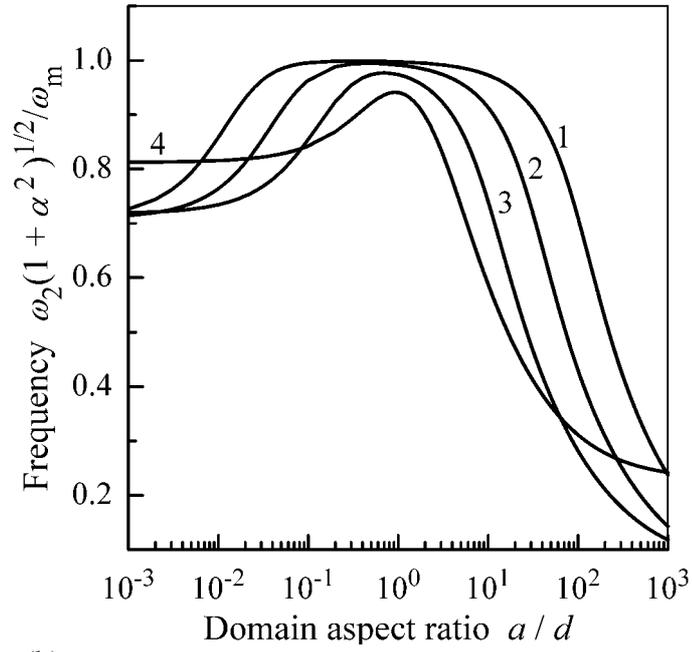

(b)

Fig. 3. Resonance frequencies $\omega_1$ (a) and $\omega_2$ (b) versus domain aspect ratio $a/d$ at $\psi = \pi/4$ and different values of quality factor $Q$: 1, $Q = 10^{-4}$; 2, $Q = 10^{-3}$; 3, $Q = 10^{-2}$; 4, $Q = 10^{-1}$.

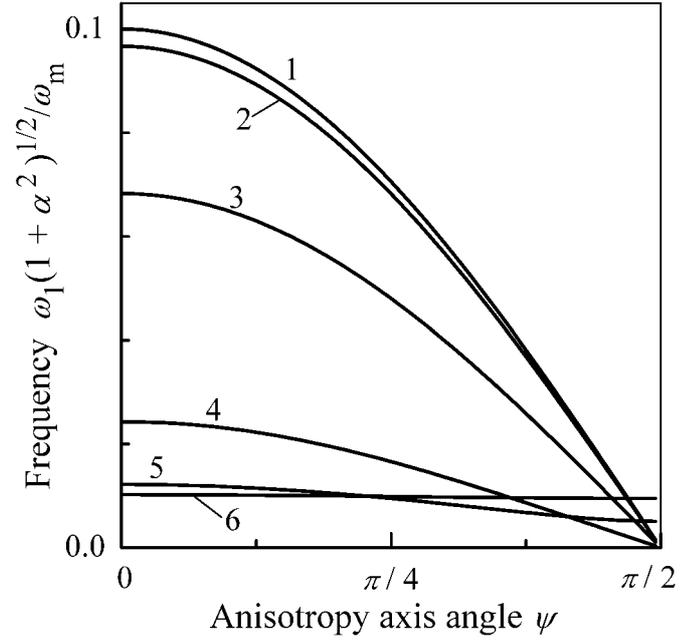

(a)

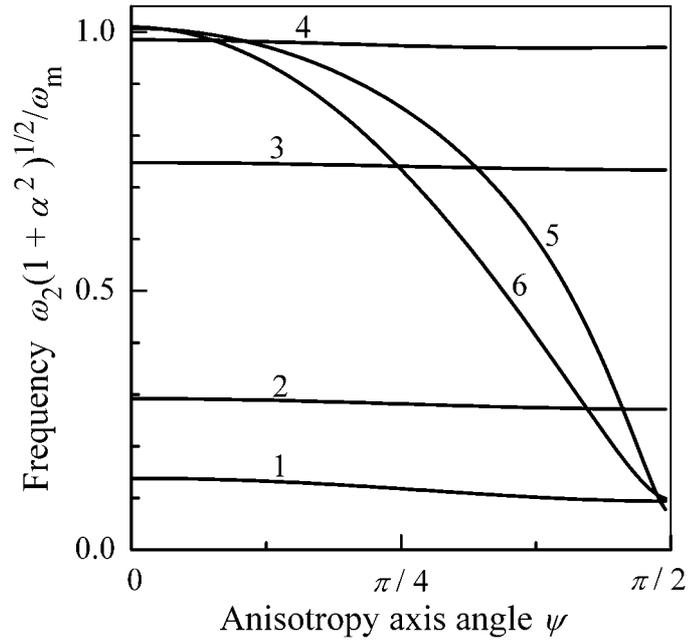

(b)

Fig. 4. Resonance frequencies $\omega_1$ (a) and $\omega_2$ (b) versus anisotropy axis angle $\psi$ at $Q=10^{-2}$ and different domain aspect ratio $a/d$: 1, $a/d=10^3$; 2, $a/d=10^2$; 3, $a/d=10$; 4, $a/d=1$; 5, $a/d=10^{-1}$; 6, $a/d=10^{-2}$.

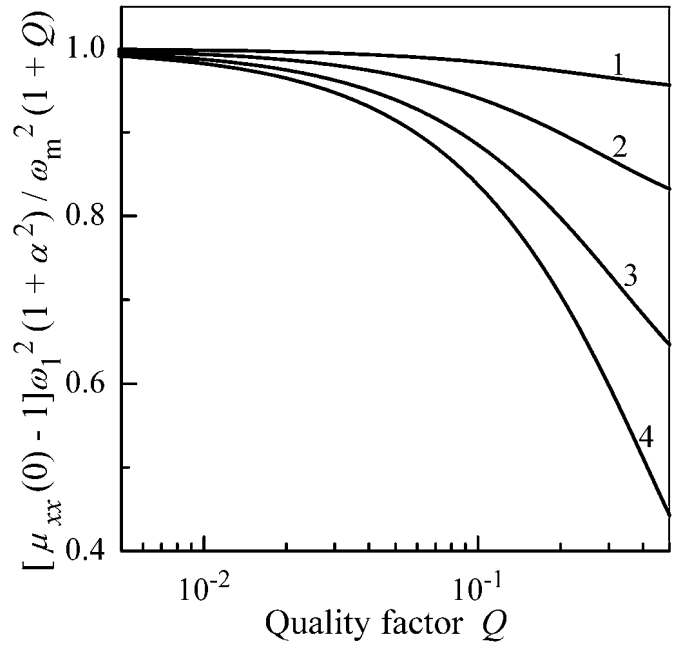

Fig. 5. Parameter $[\mu_{xx}(0)-1]\omega_1^2$ as a function of quality factor $Q$ for single-domain film at different angles $\psi$: 1, $\psi=0.1\pi$; 2, $\psi=0.2\pi$; 3, $\psi=0.3\pi$; 4, $\psi=0.4\pi$.

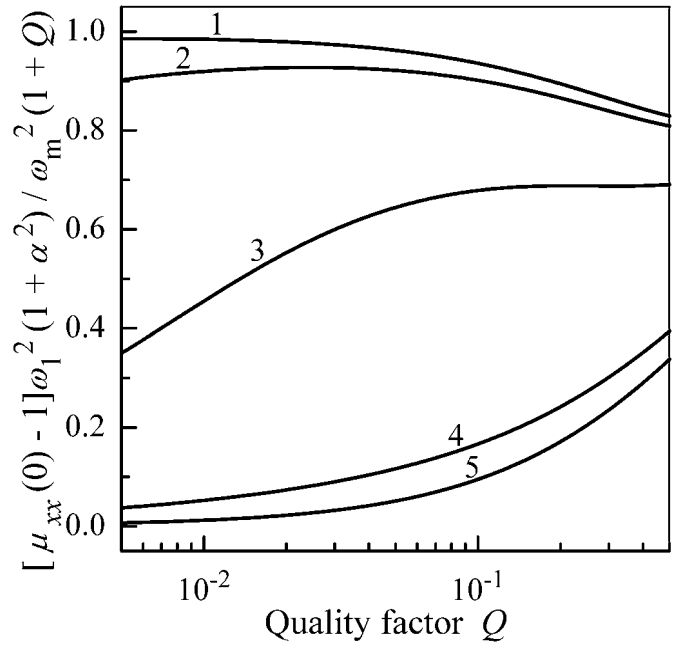

Fig. 6. Parameter $[\mu_{xx}(0)-1]\omega_1^2$ as a function of quality factor $Q$ at $\psi = 0.2\pi$ and different domain aspect ratio $a/d$: 1, $a/d = 10^3$; 2, $a/d = 10^2$; 3, $a/d = 10$; 4, $a/d = 1$; 5, $a/d = 10^{-1}$.